\documentclass[floatfix,amsmath,amssymb,superscriptaddress,a4paper,twocolumn,pra]{revtex4}
\usepackage{bm}%.......... needed for bold math
\usepackage{graphicx}%.... needed for figures
\usepackage{hyperref}%.... use for hypertext links, including those to external documents
\usepackage{color}%....... use if color is used in text

\begin{document}

\title{Coronal lines and the importance of deep core-valence correlation in Ag-like ions}

\author{Jon Grumer} 
\email{jon.grumer@teorfys.lu.se}
\affiliation{Division of Mathematical Physics, Department of Physics, Lund University, Sweden}

\author{Ruifeng Zhao}
\affiliation{The Key lab of Applied Ion Beam Physics, Ministry of Education, China}
\affiliation{Shanghai EBIT laboratory, Modern physics institute, Fudan University, Shanghai, China}

\author{Tomas Brage} 
\affiliation{Division of Mathematical Physics, Department of Physics, Lund University, Sweden}

\author{Wenxian Li}
\affiliation{The Key lab of Applied Ion Beam Physics, Ministry of Education, China}
\affiliation{Shanghai EBIT laboratory, Modern physics institute, Fudan University, Shanghai, China}

\author{Sven Huldt}
\affiliation{Lund Observatory, Lund University, Sweden}

\author{Roger Hutton} 
%\email{rhutton@fudan.edu.cn}
\affiliation{The Key lab of Applied Ion Beam Physics, Ministry of Education, China}
\affiliation{Shanghai EBIT laboratory, Modern physics institute, Fudan University, Shanghai, China}

\author{Yaming Zou}
\affiliation{The Key lab of Applied Ion Beam Physics, Ministry of Education, China}
\affiliation{Shanghai EBIT laboratory, Modern physics institute, Fudan University, Shanghai, China}

\begin{abstract}
We report on large-scale and critically evaluated {\em ab initio} MCDHF calculations of the wavelength of the "coronal", M1 transition $4f\ ^2\mathrm{F}_{5/2}^o~-~^2\mathrm{F}_{7/2}^o$ in Ag-like ions. The transition between these two fine structure levels, which makes up the ground term for $Z \ge 62$ in the isoelectronic sequence, has recently been observed in Yb$^{23+}$ and W$^{27+}$, where the latter could be of great importance for fusion plasma diagnostics. We present recommended values for all members of the sequence between $Z = 50$ and $94$, which are supported by excellent agreement with values from recent experiments. The importance of including core-valence correlation with the $n=3$ shell in the theoretical model is emphasized.  The results show close to spectroscopic accuracy for these forbidden lines.
\end{abstract}
%\pacs{32.30.Jc, 32.70.Fw, 31.15.aj}
\date{\today}

\maketitle

\section{Introduction}
Forbidden M1 transitions take place between states of the same parity. In particular such transitions among ground state levels in highly charged ions can be in the visible spectral region and have quite low transition rates. The most famous M1 transitions are the so-called \emph{coronal lines} whose origin was unknown for more than 70 years before Edl\'{e}n identified several of them as ground state M1 lines in 9-15 times ionized ions, mainly Ca, Fe and Ni in 1942 \cite{edlen1943,edlen1945}. Not for many years would it be possible to observe such lines in laboratory light sources. In 1978 Suckewer and Hinnov \cite{suckewer1978} made the first observation of an M1 transition in a fusion plasma. From the Doppler width of this line, 2665 \AA\ in Fe XX, a record temperature (at that time) of $45\times 10^6$ K was derived for the PLT tokamak. In Tokamaks, the solar corona, and in particular Electron Beam Ion Traps (EBITs) the plasma density is low enough for such lines to appear. In other terrestrial light sources for highly charged ions, e.g. sparks and laser produced plasmas, the long radiative lifetimes of the excited levels responsible for M1 transitions would lead to collisional quenching. It is interesting to note that Edl{\'e}n could not observe the M1 lines he identified in the solar corona using contemporary laboratory light sources due to density problems. His identifications were based on his established energy levels from soft x ray spectroscopy. In the same way an attempt to establish the energy difference between the Ag-like ground state $4d^{10}\,4f~^2\mathrm{F}_{5/2}$ and $^2\mathrm{F}_{7/2}^o$ levels (for $Z\ge 62$), which is the subject of the work presented here, was only done through soft x ray spectroscopy \cite{sugar1980a}. 

The actual M1 transition connecting these two levels had not been observed before our work on Ag-like W \cite{fei2012} and recently Yb \cite{zhao2014}. The lifetime of the upper level being in the millisecond range places interesting requirements on the density of the light source. EBITs with electron densities on the order of 1012 cm$^{-3}$, or less, are ideal light sources for studying such transitions. Although the electron density in EBITs is lower than Tokamak fusion plasma densities, by around 2 orders of magnitude, M1 lines have been observed in fusion devices, e.g. as mentioned above the $2s^2\, 2p^3~^2D_{5/2} - ^2D_{3/2}$ M1 decay in Fe XX \cite{suckewer1978}. Also at the National Institute for Fusion Science in Japan Morita {\it et al.} \cite{morita2013} reported the observation of M1 transitions in the visible region for highly charged tungsten ions. Previously spectra from Tokamaks made a great impact on the study of M1 and other forbidden transitions  (see \cite{martinson2001} for details). With EBITs it is possible to measure both wavelengths and lifetimes of M1 \cite{lapierre2005} and even higher order transitions in highly charged ions, for example the studies of the M3 decay in Ni-like Xe \citep{trabert2007} and later Ni-like W \cite{ralchenko2007}. 
 
M1 transitions in highly charged ions with seemingly simple ground states such as the Ag-like $4d^{10}\,4f~^2\mathrm{F}_{5/2,7/2}^o$ doublet are interesting testing grounds for theoretical methods since (a) for some ions along a sequence the M1 line will be a visible transition and therefore accessible to accurate measurement and (b) the calculation could be sensitive to  correlation from deeper bound electrons, first noted in \cite{fei2012} and further investigated in this work. Finally (c) it is also possible that the results could be useful as a test of quantum-electrodynamical effects.

In the present work we use a systematic approach to calculate the wavelength of these M1 transitions in Ag-like ions. The $ {4d^{10}\,4f~^2\mathrm{F}}$ is the ground-term for ions with $Z > 61$, while at the neutral end the ${4d^{10}\,5s~^2\mathrm{S}}$ forms the ground state. Adopting the Multiconfiguration Dirac-Hartree-Fock approach we carefully monitor the accuracy of the transition energy within different electron correlation models as a function of basis size. To further support the identification of these M1-lines, we use an isoelectronic analysis. The agreement between theory and experiment should be consistent for several ions and the trend of different atomic properties ought to behave smoothly as a function of the nuclear charge along the sequence.

Some previous isoelectronic work has been reported for theoretical work on Ag-like systems. Safronova \textit{et al.} \cite{safronova2003} used the Relativistic Many-Body Perturbation Theory (RMBPT) to study the energies of the singly excited states $4d^{10}\,\{4f,\,5s,\,5p,\,5d,\,5f,\,5g\}$ for ions between $Z=48$ and $100$. The energy structures were unfortunately only tabulated for a few selected ions at the neutral end, but the rest was made available through the more recent publication of binding energies by Ivanova \cite{ivanova2011}, from which it is possible to extract the $4f~^2\mathrm{F}^o$ fine structure energy separations. Comparison with these results along the sequence provides a reliable benchmark for the present study, especially due to the different nature, perturbative versus variational, of the two methods. Ivanova also reported a few years earlier on calculations of Ag-like ions with $Z=52$ to $86$ based on Relativistic Perturbation Theory with a Model Potential (RPTMP) \cite{ivanova2009}. Finally there is a recent, but more limited in terms of correlation, MCDHF calculation by Ding \textit{et al.} \cite{ding2012}.

The aim of the present work is to use systematic isoelectronical analyzes of electron correlation to provide solid support to the experimental identifications of the $4d^{10}\,4f~^2\mathrm{F}_{5/2}^o~-~^2\mathrm{F}_{7/2}^o$ M1 transition in Ag-like W \citep{fei2012} and Yb \cite{zhao2014} as well as future measurements in the mid- and, especially, high-$Z$ range of the sequence. These new data should also, in addition to constituting a theoretical benchmark,  be useful to the astrophysical- and fusion-plasma community.

\section{Method of calculation}
The $4f~^2\mathrm{F}_{5/2,7/2}^o$ atomic wavefunctions are determined along the Ag~I sequence using the Multiconfiguration Dirac-Hartree-Fock (MCDHF) method in the form of the most recently published version of the well-established fully relativistic {\sc Grasp2k} code \cite{jonsson2013}, originally developed by Grant and co-workers \cite{dyall1989,parpia1996}.

\subsection{Basic Multiconfiguration Dirac-Hartree-Fock theory}
The MCDHF method is outlined in detail in Grant's book \cite{grant2006} and the non-relativistic variant of the approach is covered by Froese Fisher {\it et al.} \cite{froese1997book}. Here we will only discuss the basic, and for our work most important, concepts. 

The starting point for the MCDHF theory is to define an Atomic State Function (ASF), $\left | \Gamma J^\pi \right \rangle$, as a linear combination of Configuration State Functions (CSFs), $\left | \gamma_i J^\pi \right \rangle$;
\begin{equation}
\left | \Gamma J^\pi \right \rangle = \sum_i c_i \left | \gamma_i J^\pi \right \rangle ~,
\end{equation}
where  $\gamma_i$ are labels to uniquely define the CSFs and $c_i$ are expansion coefficients. The $\Gamma$ is usually chosen as the $\gamma_i$ of the CSF with maximum weight $c_i^2$. 
The CSFs are in turn anti-symmetrized products of single-electron Dirac orbitals coupled to Eigenfunctions of the total angular momentum ($J ^2$ and $J_z$) and parity ($\pi$) operators.
Without going into any details the MCDHF approach is essentially a multireference self-consistent field method based on the many-body Dirac-Coulomb Hamiltonian, expressed as
\begin{equation} \label{eq:dc}
\mathcal{H}_{DC} = \sum_i^N h_D(\bm r_i) + \sum_{i>j}^N1/r_{ij}~,
\end{equation}
in Hartree atomic units. Here $h_D$ is the standard one-particle Dirac Hamiltonian and the second sum represents the instantaneous, inter-electronic Coulomb interaction. 

The CSF basis expansion is generated in an Active Space (AS) approach in which a limited number of Dirac orbitals are divided into an inactive and active set. The CSF expansion is then formed through single (S), double (D), triple (T) etc. substitutions from a set of predefined important CSFs, the multireference (MR) set, to the active set of orbitals. A calculation on the MR set builds the zero order wavefunction. Orbitals of closed shells in the MR set are typically defined as inactive and therefore not a part of the active set.

The set of Dirac orbitals and mixing coefficients are optimized to self-consistency in the MCDHF procedure, followed by a relativistic configuration interaction (RCI) calculation in order to include the Breit interaction and leading QED effects. The Breit interaction is evaluated in the low frequency limit of the exchanged virtual photon. The contribution from vacuum polarization is included to second- (Uehling) and fourth-order (K\"all\'en-Sabry) \cite{fullerton1976}) and the self-energy is evaluated in the hydrogenic approximation with reference values from \cite{mohr1983}).

The computational accuracy is essentially determined by whether the necessarily finite set of CSFs is effectively complete for the atomic states under investigation. This is dependent on the choice of included CSFs, but also on the optimization of and constraints on the Dirac orbitals. In practice the accuracy of the method is evaluated through careful convergence studies of atomic properties as a function of different correlation models and CSF-expansions within these models. The latter is defined by the size of the active set of correlation orbitals. In {\sc Grasp2k} the calculations are performed in a layer-by-layer scheme, in which the AS of CSFs is enlarged systematically. The orbitals belonging to previous layers, defined by e.g. their principal quantum number $n$, are kept fixed in the variational procedure and only the new ones are optimized.

\subsection{Correlation models}
Two different computational models are presented in this work. The first (labeled SCV) is designed to provide information about important correlation contributions. Based on the experience gained from this calculation it is possible to design a large-scale model (labeled FCV) with the goal of reaching high accuracy enough for what we could label as {\em single-line spectroscopy}.

Both models use $[1s^2\,2s^2\,...\,4d^{10}\,4f]_{5/2,7/2}^o$ as the MR set, i.e. two separate CSFs build the $J=5/2$ and $7/2$ symmetry blocks.  These CSFs are constructed from a common set of orbitals, optimized on a linear combination of the energies of the lowest state of each block (extended optimal level). In this work the Dirac-Fock method is defined as the case when the CSF expansion only includes the MR set. The orbitals obtained in the initial DF step are then kept frozen throughout the remaining procedure. To include correlation, the basis set is enlarged through substitutions from this reference configuration to a systematically increased set of CSFs. 

\subsubsection*{A Separate Core-Valence (SCV) correlation model}
In order to obtain an {\it ab initio} transition energy of close to spectroscopic accuracy, we need a detailed investigation of the correlation between valence and core electrons, or core-valence (CV) correlation. In the MCDHF scheme, CV correlation is represented by CSFs obtained from simultaneous replacements of one core and one valence electron of the CSFs in the MR set, with those in the active set of orbitals. In the special case of a singly occupied valence subshell, such as in the $4f$ configuration of Ag-like ions, the inclusion of CV correlation will in general increase the binding energy of this electron as compared to a fixed core calculation. The orbital of the single valence electron will therefore in many cases be contracted, which might have a large impact on different atomic properties. 

CV correlation is often thought of as the MCDHF representation of core polarization. This is however a too simplistic interpretation. CV correlation does in general also include radial correlation through CSFs which only differ in the principal quantum number $n$ from the reference CSFs. It is also clear that true core polarization should be evaluated by comparisons with results using core orbitals optimized on the bare core only, as the $4f$-electron polarizes the core, and not with the DF results of the $|\ldots 4d^{10}\,4f~^2\mathrm{F}^o\rangle$ states as we do here. We will therefore refer to core-valence correlation rather than core polarization.

Turning to the $4d^{10}\, 4f$ states of Ag-like ions, it was recently shown for Ag-like W \cite{fei2012} that a major part of the contribution from core-valence correlation to the fine structure separation $4f~^2\mathrm{F}_{5/2,7/2}^o$ is due to interaction with the $3d$ subshell. This is maybe counter-intuitive as one would expect that the largest contributions should come from the outermost core subshells, i.e. $4d$ in this case. We will investigate this further along the Ag-like sequence.

Defining the singly occupied $4f$ subshell as the only valence shell implies that there is no valence-valence (VV) correlation. This allows for separate studies of the energy contributions from interactions between the valence electron and the different core subshells, one subshell at a time. Such a Separate Core-Valence (SCV) study should provide valuable information about electron correlation, usable when designing a large-scale model including possible "interference" effects between contributions from different core subshells.

To be more specific, the SCV-calculations proceed with separate calculations for each subshell contribution, including only CSFs with one hole in a distinct core subshell. As an example, if we include only CSFs of the form $1s^2\,2s^2\ldots 3p^5\ldots 4d^{10}\,nl\,n'l'$, where an electron from the $3p$ core subshell is allowed to be excited together with the $4f$ valence electron, we include CV correlation with $3p$. We aim in each calculation for converged results of the $^2\mathrm{F}$ energy separation, as a function of the maximum $n$ and $l$ of the orbitals included in the active set. Taking the difference of the converged and the DF energy separation gives an estimate of the energy contribution due to CV correlation with the chosen core subshell ($3p$ in our example). Adding all these SCV contributions to the DF energy value, gives an estimate of the total fine structure separation. It should be clear that this approach only is applicable to systems with a single valence electron since it otherwise is impossible to separate the VV and CV contributions.

\subsubsection*{Full Core-Valence (FCV) correlation model}
With the results from the SCV model at hand, it is feasible to design a large-scale model in which CV correlation with all subshells (except $1s$) is included simultaneously in the MCDHF procedure. This will be referred to as the Full Core-Valence (FCV) model.

This model contains CSFs generated from simultaneous substitutions of at most one electron from any subshell down to $n=2$, together with the $4f$ valence electron of the reference configuration to the active set of orbitals. The $1s$ subshell is kept closed as it proved having a negligible effect on the $^2\mathrm{F}^o$ energy separation. Furthermore the CSFs of the $4d^8\, 4f^3$ configuration (the most important CSFs in the $n=4$ complex in addition to $4d^ {10}\, 4f$) are also added. The orbital set is increased up to $n=10$ and $l=6$ ($i$-orbitals), which corresponds to a maximum of $39\, 230$ CFSs in the $J=5/2$ block and $43\, 857$ CSFs in the $J=7/2$ block. These seem to be quite reasonably sized basis sets at first glance, but the calculation still takes a few weeks to run per charge state (with the serial version of the codes on modern 3.7 GHz Intel Xeon-based computers) due to comparatively dense energy matrices.

\section{Results and Discussion}
We start this section with a discussion of the experiences gained from the smaller SCV model. This is followed by results from the large-scale FCV calculation, together with comparisons with other recent theoretical and experimental results. 
There is a special focus on the differences in the amount of core-valence correlation included in the different models and the impact of this on the final fine structure separations for different members of the isoelectronic sequence. In the  final section we present rates for the magnetic-dipole transitions.

\subsection{The impact of core-valence correlation}
\begin{figure*}[ht]
\includegraphics[width=0.65\textwidth, angle=0]{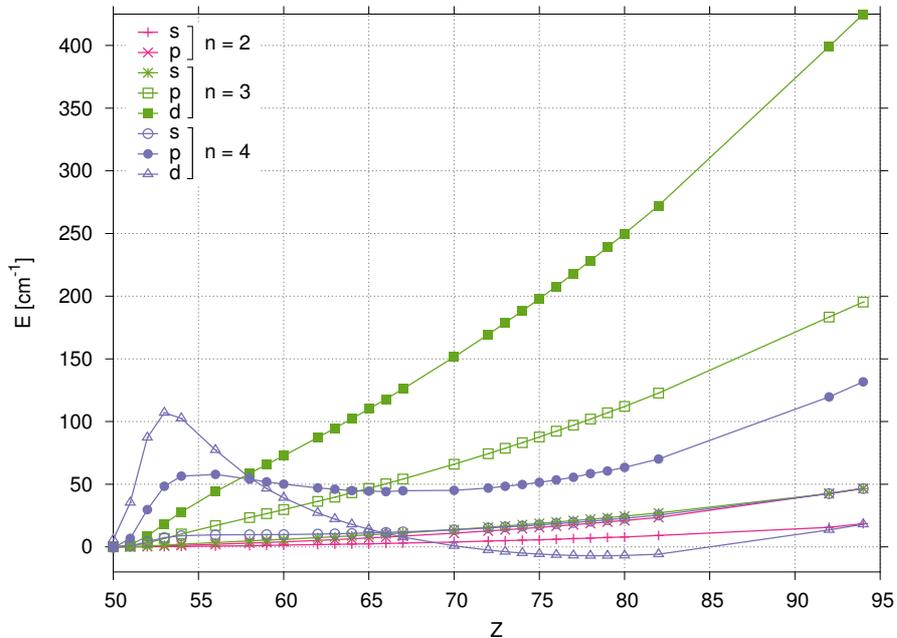}
\caption{\label{fig:scv_contrib} Absolute contributions from core-valence correlation with different core subshells relative to Dirac-Fock energies ($E=E_\mathrm{tot}^\mathrm{SCV}(nl)-E_\mathrm{DF}$, the SCV model is explained in the text) to the $4f~^2\mathrm{F}^o$ fine structure energy separation of Ag-like ions with nuclear charges $50 \le Z \le 94$. This clearly shows the dominating behavior of core-valence correlation with $3d$ rather than $4d$ in the mid- and high-$Z$ regime. Note that these energy contributions are presented as absolute numbers and not as a fraction of the total fine structure energy separation, to compared with Tab. \ref{tab:fcv} or Fig. \ref{fig:fcv} where the total energies are given.}
\end{figure*}

The SCV calculation reveals interesting trends of the effect of core-valence correlation on the fine structure separation along the isoelectronic sequence as can be seen in Fig. \ref{fig:scv_contrib}. In this plot the energy contribution due to CV correlation with all core subshells down to $2s$ is presented for ions between $Z=50$ and $92$. For the ions at the neutral end it is clear that the CV contribution from $4d$ is large as would be expected. However, for $Z \ge 58$ the major contribution is due to $3d$ and it becomes increasingly dominant as $Z$ increases, followed by CV correlation with $3p$ and $4p$. For $Z=94$ the correlation with $3d$ makes up $46\%$ of the total contribution. It is also interesting to note that the impact of correlation with $4d$ becomes almost negligible for $Z\ge 70$ and actually gives a negative contribution for $70<Z<85$. The fact that CV correlation with deeper core shells becomes an important factor in the calculation of this fine structure separation was first noted for Ag-like W ($Z=74$) \cite{fei2012} where correlation with $3d$ contributes with $51\%$ of the total value and the whole $n=3$ shell $78\%$.

Judging from the results of this initial investigation, one can conclude that core-valence correlation with essentially all core subshells is of importance to the fine structure separation when aiming for high accuracy. In the low-$Z$ regime it's clear that interactions with the $n=4$ subshells are crucial, replaced by $3p$ and foremost $3d$ for higher members of the sequence. 

\subsection{The \texorpdfstring{$4f~^2\mathrm{F}^o$} \ \ fine structure separation from the FCV model}

\begin{figure}[ht]
\includegraphics[width=0.4\textwidth, angle=-90]{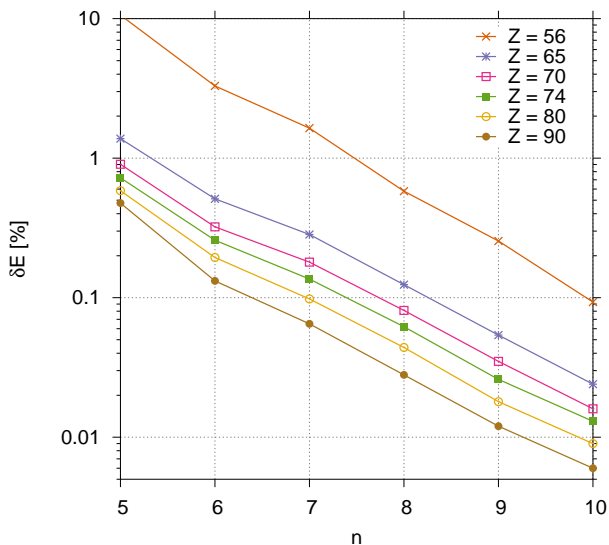}
\caption{\label{fig:conv} Relative convergence of the $^2\mathrm{F}^o$ energy separation as the size of the active set of orbitals is increased in a layer-by-layer scheme, denoted by the principal quantum number $n$. $\delta E$ is the difference in percentages of energy from the previous correlation layer.}
\end{figure}

\begin{table*}
\caption{\label{tab:fcv}The fine structure separation of $4f\ ^2\mathrm{F}_{5/2,7/2}^o$ from the Full Core-Valence (FCV) calculation (see text for details). The first column shows the atomic number, $Z$ and the second ($E_\mathrm{DF}^\mathrm{DC}$) gives the energy separations resulting from single-CSF calculations, here referred to as Dirac-Fock (DF), based on the Dirac-Coulomb (DC) Hamiltonian. The third column ($\delta E_\mathrm{B}$) presents additional energy contributions due to the Breit interaction in the low-frequency limit (B). The fourth ($\delta E_\mathrm{QED}$) presents the total contribution from self-energy and vacuum polarization (QED) corrections. The fifth column ($\delta E_\mathrm{corr}$) shows how big the influence of correlation is in the DC+B+QED scheme. In the sixth column ($E_\mathrm{tot}$) the total energy separations (including Breit, QED and correlation) are presented. All energies are given in cm$^{-1}$ and a negative total energy value corresponds to an inverted fine structure (i.e. the $J=7/2$ level having lowest energy).}
%\footnotesize
\begin{tabular*}{\textwidth}{@{\extracolsep{\fill}}crrrrrccrrrrr}
\hline \hline
$Z$      &$E_\mathrm{DF}^\mathrm{DC}$ &$+\delta E_\mathrm{B}$  &$+\delta E_\mathrm{QED}$  &$+\delta E_\mathrm{corr}$ &$= E_\mathrm{tot}$ &  &
$Z$      &$E_\mathrm{DF}^\mathrm{DC}$ &$+\delta E_\mathrm{B}$  &$+\delta E_\mathrm{QED}$  &$+\delta E_\mathrm{corr}$ &$= E_\mathrm{tot}$ \\ \cline{1-6}\cline{8-13}
$50$     &$-88  $ &$-3   $ &$0 $    &$6  $     &$-85  $   &     &$73$     &$27826 $    &$-1400$   &$20 $     &$341$     &$26786 $ \\
$51$     &$-182 $ &$-13  $ &$0 $    &$59 $     &$-136 $   &     &$74$     &$30750 $    &$-1510$   &$22 $     &$358$     &$29619 $ \\
$52$     &$-161 $ &$-37  $ &$0 $    &$157$     &$-41  $   &     &$75$     &$33876 $    &$-1626$   &$25 $     &$376$     &$32651 $ \\
$53$     &$62   $ &$-70  $ &$0 $    &$213$     &$205  $   &     &$76$     &$37215 $    &$-1747$   &$27 $     &$394$     &$35890 $ \\
$54$     &$441  $ &$-106 $ &$0 $    &$229$     &$564  $   &     &$77$     &$40774 $    &$-1873$   &$31 $     &$414$     &$39346 $ \\
$55$     &$925  $ &$-145 $ &$1 $    &$229$     &$1010 $   &     &$78$     &$44564 $    &$-2005$   &$34 $     &$436$     &$43028 $ \\
$56$     &$1495 $ &$-186 $ &$1 $    &$225$     &$1535 $   &     &$79$     &$48592 $    &$-2143$   &$37 $     &$458$     &$46945 $ \\
$57$     &$2145 $ &$-229 $ &$1 $    &$221$     &$2139 $   &     &$80$     &$52869 $    &$-2286$   &$41 $     &$481$     &$51106 $ \\
$58$     &$2877 $ &$-274 $ &$2 $    &$218$     &$2823 $   &     &$81$     &$57405 $    &$-2435$   &$45 $     &$506$     &$55521 $ \\
$59$     &$3696 $ &$-323 $ &$2 $    &$217$     &$3592 $   &     &$82$     &$62209 $    &$-2590$   &$50 $     &$531$     &$60199 $ \\
$60$     &$4606 $ &$-374 $ &$3 $    &$217$     &$4451 $   &     &$83$     &$67290 $    &$-2752$   &$55 $     &$557$     &$65151 $ \\
$61$     &$5613 $ &$-429 $ &$3 $    &$219$     &$5406 $   &     &$84$     &$72660 $    &$-2919$   &$60 $     &$585$     &$70385 $ \\
$62$     &$6724 $ &$-488 $ &$4 $    &$223$     &$6463 $   &     &$85$     &$78329 $    &$-3093$   &$65 $     &$613$     &$75914 $ \\
$63$     &$7946 $ &$-550 $ &$5 $    &$227$     &$7628 $   &     &$86$     &$84307 $    &$-3274$   &$71 $     &$643$     &$81747 $ \\
$64$     &$9286 $ &$-616 $ &$5 $    &$234$     &$8909 $   &     &$87$     &$90604 $    &$-3461$   &$77 $     &$674$     &$87894 $ \\
$65$     &$10749$ &$-686 $ &$6 $    &$241$     &$10311$   &     &$88$     &$97232 $    &$-3656$   &$84 $     &$706$     &$94366 $ \\
$66$     &$12344$ &$-760 $ &$8 $    &$250$     &$11842$   &     &$89$     &$104201$    &$-3857$   &$91 $     &$740$     &$101175$ \\
$67$     &$14078$ &$-838 $ &$9 $    &$260$     &$13509$   &     &$90$     &$111524$    &$-4065$   &$98 $     &$774$     &$108331$ \\
$68$     &$15958$ &$-920 $ &$10$    &$271$     &$15320$   &     &$91$     &$119210$    &$-4280$   &$106$     &$810$     &$115845$ \\
$69$     &$17992$ &$-1006$ &$12$    &$283$     &$17280$   &     &$92$     &$127272$    &$-4503$   &$114$     &$847$     &$123729$ \\
$70$     &$20188$ &$-1098$ &$13$    &$295$     &$19399$   &     &$93$     &$135721$    &$-4734$   &$122$     &$886$     &$131995$ \\
$71$     &$22554$ &$-1193$ &$15$    &$310$     &$21685$   &     &$94$     &$144569$    &$-4972$   &$132$     &$925$     &$140654$ \\
$72$     &$25097$ &$-1294$ &$17$    &$325$     &$24145$   &     &         &            &            &          &         \\
\hline \hline
\end{tabular*}
\end{table*}

\begin{figure}[ht]
\includegraphics[width=0.33\textwidth, angle=-90]{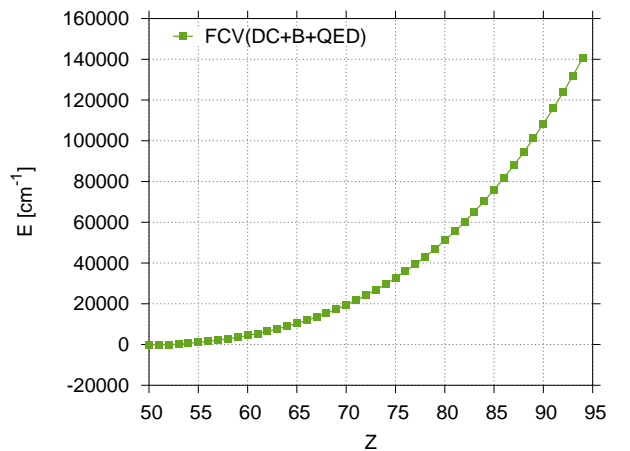}
\caption{\label{fig:fcv} Energies of the $4d^{10}\, 4f~^2\mathrm{F}^o$ fine structure separation from our FCV calculation along the Ag-like isoelectronic sequence.}
\end{figure}

In this section we present results from $^2\mathrm{F}^o$ energy separations along the Ag-like isoelectronic sequence from the large-scale FCV model in which core-valence correlation is included with all subshells except $1s$. The model has been carefully evaluated in terms of convergence of the energy separation of $^2\mathrm{F}^o$ with respect to the size of the active set of Dirac orbitals. Within the boundaries of this model it can be seen from Fig. \ref{fig:conv},  which shows the convergence trend (difference in energy from previous correlation layer as a fraction of the total fine structure) in a logarithmic scale, that the fine structure separation has been converged to close to $0.1\%$ for $Z=56$ and to $ 0.006\%$ for $Z=90$.

Resulting energies for all ions in the Ag-like isoelectronic sequence with nuclear charges $50 \le Z \le 94$ are presented in Tab. \ref{tab:fcv} and Fig. \ref{fig:fcv}. In the second column we give the Dirac-Fock energies from the MCDHF (Dirac-Coulomb) approach, where a negative value corresponds to an inverted fine structure. The third column contains energy contributions due to the frequency independent Breit interaction, and the fourth shows leading QED effects. The impact of electron correlation in the regime of the Dirac-Coulomb-Breit-QED Hamiltonian (calculated by taking the difference of the DF and the converged FCV results) is presented in the fifth column. Finally the sixth column and Fig. \ref{fig:fcv} give the total energy separation including all the above mentioned contributions. It is clear that correlation is the dominating correction to DF (Dirac-Coulomb) in the low-$Z$ regime, replaced by the Breit interaction for $Z \ge 57$. The energy shift due to the QED corrections are comparatively small along the whole sequence.

\subsection{Comparison with experiment and other theory}

% Make rows in table slightly tighter to fit in one page
%{ \renewcommand{\arraystretch}{0.95} 

\begin{table*}[ht!]
\caption{\label{tab:levs-comp} Comparison of the $4f~^2\mathrm{F}_{5/2,7/2}^o$ energy separation obtained from the large-scale FCV model ($E_\mathrm{tot}^\mathrm{FCV}$) with experiment ($E_\mathrm{exp}$) (corresponding source(s) are given in the fourth column), and other available theory ($E_\mathrm{RMBPT}$ \cite{safronova2003}, $E_\mathrm{MCDHF}$ \cite{ding2012} and $E_\mathrm{RPTMP}$ \cite{ivanova2009,ivanova2011}). All energies are given in cm$^{-1}$ and the differences are presented relative to the FCV values of this work in absolute numbers $\delta_E$ and in percentages $\delta_\%$.
}
%\footnotesize
\begin{tabular*}{\textwidth}{@{\extracolsep{\fill}}rrrllrrrrrrrrrrr}
\hline \hline
Z  &\multicolumn{1}{c}{$E_\mathrm{tot}^\mathrm{FCV}$}&\multicolumn{2}{c}{$E_\mathrm{exp}$} &Source  &$\delta_E$ &$\delta_\%$   &$E_\mathrm{RMBPT}$  &$\delta_E$ &$\delta_\%$   &$E_\mathrm{MCDHF}$ &$\delta_E$ &$\delta_\%$   &$E_\mathrm{RPTMP}$ & $\delta_E$ &$\delta_\%$ \\ 
\hline
50 &$-$85   & $-$60 &           & \cite{moore1958,NIST}     &$-$24 &  40$\%$ &$-$76               &$-$9  &    12$\%$ &$-$71  &$-$14  &    20$\%$     &      &  &                  \\ 
51 &$-$136  &       &           &                           &      &         &$-$162$^\mathrm{b}$ &26    & $-$16$\%$ &$-$121 &$-$15  &    12$\%$     &      &  &                  \\
52 &$-$41   &       &           &                           &      &         &                    &      &           &$-$118 &77     & $-$66$\%$     &      &  &                  \\
53 &205     & 200   &           & \cite{kaufman1981a}       &   5  & 2.7$\%$ &184                 &21    &    12$\%$ &71     &134    &   189$\%$     &      &  &                  \\
54 &564     & 550   &           & \cite{larsson1995,NIST}   &   14 & 2.5$\%$ &542                 &22    &    4.0$\%$&411    &153    &    37$\%$     &      &  &                  \\
55 &1010    & 987   &           & \cite{tauheed2005,NIST}   &   23 & 2.3$\%$ &                    &      &           &854    &156    &    18$\%$     &      &  &                  \\
56 &1535    & 1516  &           & \cite{churilov2000,NIST}  &   19 & 1.3$\%$ &                    &      &           &1380   &155    &    11$\%$     &      &  &                  \\
57 &2139    & 2160  &           & \cite{kaufman1981a}       &$-$21 &$-$1.0$\%$&2123               &16    &    0.7$\%$&1984   &155    &   7.8$\%$     &      &  &                  \\
58 &2823    & 2784  &           & \cite{sugar1981b}         &   39 & 1.4 $\%$&2810                &13    &    0.4$\%$&2672   &151    &   5.6$\%$     &      &  &                  \\
59 &3592    & 3577  &           & \cite{sugar1981b}         &   15 & 0.4 $\%$&                    &      &           &3442   &150    &   4.3$\%$     &      &  &                  \\ \hline
60 &4451    & 4430  &           & \cite{sugar1981b}         &   21 & 0.5 $\%$&                    &      &           &4302   &149    &   3.5$\%$     &      &  &                  \\
61 &5406    &       &           &                           &      &         &5476$^\mathrm{c}$   &$-$70 & $-$1.3$\%$&5253   &153    &   2.9$\%$     &5272  &134      &   2.5$\%$ \\
62 &6463    & 6555  &           & \cite{sugar1981b}         &$-$92 &$-$1.4$\%$&6533$^\mathrm{c}$  &$-$70 & $-$1.1$\%$&6301   &162    &   2.6$\%$     &6353  &110      &   1.7$\%$ \\
63 &7628    & 7521  & $\pm$ 62  & \cite{sugar1981b}         &   107&  1.4$\%$&7697$^\mathrm{c}$   &$-$69 & $-$0.9$\%$&7444   &184    &   2.5$\%$     &7504  &124      &   1.7$\%$ \\
64 &8909    & 8900  &           & \cite{sugar1981b}$^\mathrm{a}$&9 &         &8977$^\mathrm{c}$   &$-$68 & $-$0.8$\%$&8685   &224    &   2.6$\%$     &8800  &109      &   1.2$\%$ \\
65 &10311   & 10280 &           & \cite{sugar1981b}         &   31 &  0.3$\%$&10378$^\mathrm{c}$  &$-$67 & $-$0.6$\%$&10033  &278    &   2.8$\%$     &10225 &86       &   0.8$\%$ \\
66 &11842   & 11770 & $\pm$ 131 & \cite{sugar1981b}         &   72 &  0.6$\%$&11908$^\mathrm{c}$  &$-$66 & $-$0.6$\%$&11512  &330    &   2.9$\%$     &11870 &$-$28    &   0.2$\%$ \\
67 &13509   & 13500 &           & \cite{sugar1981b}$^\mathrm{a}$&9 &  0.1$\%$&13573$^\mathrm{c}$  &$-$64 & $-$0.5$\%$&13140  &369    &   2.8$\%$     &13486 &23       &   0.2$\%$ \\
68 &15320   &       &           &                           &      &         &15383$^\mathrm{c}$  &$-$63 & $-$0.4$\%$&14926  &394    &   2.6$\%$     &15526 &$-$206   &$-$1.3$\%$ \\
69 &17280   &       &           &                           &      &         &17341$^\mathrm{c}$  &$-$61 & $-$0.3$\%$&16871  &409    &   2.4$\%$     &17695 &$-$415   &$-$2.3$\%$ \\ \hline
70 &19399   & 19383 & $\pm$ 8   & \cite{zhao2014}           &   16 &  0.1$\%$&19459$^\mathrm{c}$  &$-$60 & $-$0.3$\%$&18979  &420    &   2.2$\%$     &19848 &$-$449   &$-$2.3$\%$ \\
71 &21685   &       &           &                           &      &         &21741$^\mathrm{c}$  &$-$56 & $-$0.3$\%$&21254  &431    &   2.0$\%$     &22465 &$-$780   &$-$3.5$\%$ \\
72 &24145   &       &           &                           &      &         &24198$^\mathrm{c}$  &$-$53 & $-$0.2$\%$&23702  &443    &   1.9$\%$     &25285 &$-$1140  &$-$4.5$\%$ \\
73 &26786   &       &           &                           &      &         &26838$^\mathrm{c}$  &$-$52 & $-$0.2$\%$&26331  &455    &   1.7$\%$     &28350 &$-$1564  &$-$5.5$\%$ \\
74 &29619   & 29600 & $\pm$ 2   & \cite{fei2012}            &   19 &  0.1$\%$&29668$^\mathrm{c}$  &$-$49 & $-$0.2$\%$&29151  &468    &   1.6$\%$     &31769 &$-$2150  &$-$6.8$\%$ \\
75 &32651   &       &           &                  &      &         &32696$^\mathrm{c}$  &$-$45 & $-$0.1$\%$&32167  &484    &   1.5$\%$     &35494 &$-$2843  &$-$8.0$\%$ \\
76 &35890   &       &           &                  &      &         &35932$^\mathrm{c}$  &$-$42 & $-$0.1$\%$&35390  &500    &   1.4$\%$     &39491 &$-$3601  &$-$9.1$\%$ \\
77 &39346   &       &           &                  &      &         &39385$^\mathrm{c}$  &$-$39 & $-$0.1$\%$&38828  &518    &   1.3$\%$     &43765 &$-$4419  &$-$10$\%$  \\
78 &43028   &       &           &                  &      &         &43063$^\mathrm{c}$  &$-$35 & $-$0.1$\%$&42491  &537    &   1.3$\%$     &48320 &$-$5292  &$-$11$\%$  \\
79 &46945   &       &           &                  &      &         &46976$^\mathrm{c}$  &$-$31 & $-$0.1$\%$&46387  &558    &   1.2$\%$     &53411 &$-$6466  &$-$12$\%$  \\ \hline
80 &51106   &       &           &                  &      &         &51133$^\mathrm{c}$  &$-$27 & $-$0.1$\%$&50527  &579    &   1.1$\%$     &58754 &$-$7648  &$-$13$\%$  \\
81 &55521   &       &           &                  &      &         &55542$^\mathrm{c}$  &$-$21 &   0.0 $\%$&54918  &603    &   1.1$\%$     &64649 &$-$9128  &$-$14$\%$  \\
82 &60199   &       &           &                  &      &         &60216$^\mathrm{c}$  &$-$17 &   0.0 $\%$&59573  &626    &   1.1$\%$     &70791 &$-$10592 &$-$15$\%$  \\
83 &65151   &       &           &                  &      &         &65162$^\mathrm{c}$  &$-$11 &   0.0 $\%$&64499  &652    &   1.0$\%$     &77338 &$-$12187 &$-$16$\%$  \\
84 &70385   &       &           &                  &      &         &70391$^\mathrm{c}$  &$-$6  &   0.0 $\%$&69708  &677    &   1.0$\%$     &84066 &$-$13681 &$-$16$\%$  \\
85 &75914   &       &           &                  &      &         &75914$^\mathrm{c}$  &0     &   0.0 $\%$&75209  &705    &   0.9$\%$     &91483 &$-$15569 &$-$17$\%$  \\
86 &81747   &       &           &                  &      &         &81741$^\mathrm{c}$  &6     &   0.0 $\%$&81012  &735    &   0.9$\%$     &99145 &$-$17398 &$-$18$\%$  \\
87 &87894   &       &           &                  &      &         &                    &      &           &87129  &765    &   0.9$\%$     &      &         &           \\
88 &94366   &       &           &                  &      &         &                    &      &           &93571  &795    &   0.8$\%$     &      &         &           \\
89 &101175  &       &           &                  &      &         &                    &      &           &100347 &828    &   0.8$\%$     &      &         &           \\ \hline
90 &108331  &       &           &                  &      &         &                    &      &           &107469 &862    &   0.8$\%$     &      &         &           \\
91 &115845  &       &           &                  &      &         &                    &      &           &114944 &901    &   0.8$\%$     &      &         &           \\
92 &123729  &       &           &                  &      &         &                    &      &           &122792 &937    &   0.8$\%$     &      &         &           \\
93 &131995  &       &           &                  &      &         &                    &      &           &       &       &               &      &         &           \\
94 &140654  &       &           &                  &      &         &                    &      &           &       &       &               &      &         &           \\
\hline \hline 
\multicolumn{7}{l}{$^\mathrm{a}$ based on interpolated values of $4f~^2\mathrm{F}_{7/2}^o$ \cite{sugar1981b}} & 
\multicolumn{4}{l}{$^\mathrm{b}$ from Tab. 8  in Ref. \cite{ivanova2009}} & 
\multicolumn{5}{l}{$^\mathrm{c}$ from Tab. 11 in Ref. \cite{ivanova2011}} \\
\multicolumn{15}{c}{~} \\
\end{tabular*}
\end{table*}
%} % end arraystretch

\begin{figure*}[ht]
\includegraphics[width=0.45\textwidth, angle=-90]{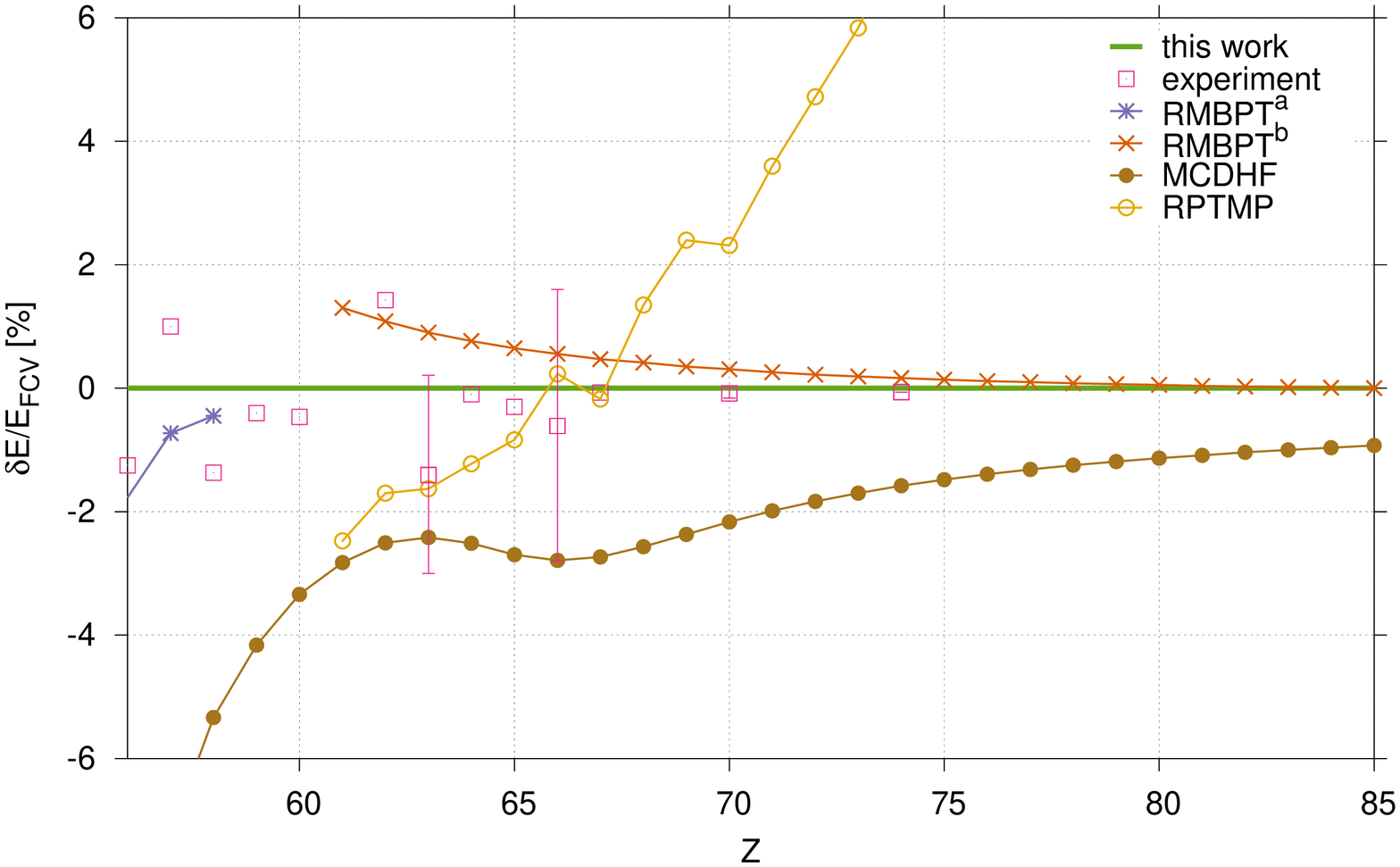}
\caption{\label{fig:obscalc} 
Experimental (see Tab. \ref{tab:levs-comp} for sources) and other theoretical (RMBPT$^a$ \cite{safronova2003}, RMBPT$^b$ \cite{safronova2003} (from \cite{ivanova2011}), MCDHF \cite{ding2012} and RPTMP \cite{ivanova2009,ivanova2011}) energies presented for an interesting sub-range ($56 \ge Z \ge 85  $) of the Ag-like isoelectronic sequence. The energy separations of the different sources are plotted as fractional differences relative to the FCV calculation $\delta E/E_\mathrm{FCV}^\mathrm{tot}=(E_\mathrm{method}-E_\mathrm{FCV}^\mathrm{tot})/E_\mathrm{FCV}^\mathrm{tot}$ in percentages. Error bars of four experimental data points were available and are included in the plot.}
\end{figure*}

The fine structure energy separations from experimental and other theoretical results are compared to our FCV values in  Tab. \ref{tab:levs-comp} and plotted as differences to our FCV results in Fig. \ref{fig:obscalc}. As shown in Fig. \ref{fig:fcv}, the trend of the fine structure splitting along the sequence should be smooth, in the absence of level crossing or other effects. We should therefore expect a smilar behavior for other methods and thereby also for the difference between different sets of results.

The agreement of our results with most of the experimental data points is in general very good, with the largest deviations between
$Z=62$ to $66$, where the $4d^{10}\, 4f~^2\mathrm{F}^o$ term becomes ground the state ($Z=62$). The experimental values do show, however, an irregular isoelectronic trend for low-Z ions, which do warrant further investigations. For higher nuclear charges there are excellent agreement (less than $0.1\%$) with the two most recent experimental 
results: $19\, 383$ cm$^{-1}$ for Yb ($Z=70$) \cite{zhao2014} and $29\, 600$ cm$^{-1}$ for W ($Z=74$) 
\cite{fei2012}. 

There is also a good agreement between our results and the relativistic many-body perturbation theory 
(RMBPT) calculation, especially in the beginning and high-end of the sequence. More importantly in the two ends, the difference between the two data sets behaves in smooth way, except for the leap between $Z=57$ and $61$. It is important to note, however, that the 
RMBPT values are collected from two sets of data, presented in two different publications ($Z\le 57$ from 
Safronova {\it et al.}  \cite{safronova2003} and $Z\ge 61 $ from Ivanova \cite{ivanova2011}). The leap in energy might therefore be due to some inconsistency between the two papers or in the model. Another  possibility is close degeneracy caused by level crossings as the $4f$ configuration becomes the ground state, which could be difficult to represent 
in a perturbative approach. 

 The difference between our results and the earlier MCDHF calculation \cite{ding2012} is most likely 
explained by their exclusion of core-valence correlation with other subshells than $4d$. From the earlier discussion about the SCV investigation, presented in Fig. \ref{fig:scv_contrib}, it was made clear the $n=3$ shell contributes around $78\%$ of the total amount of core-valence correlation whereas the $4d$ subshell barely contributes at all for this $Z$. It is however hard to understand the irregular isoelectronic behavior of their results.

Finally it is clear that there is a large inconsistency between the RPTMP results \cite{ivanova2009} and all other methods presented here, since the isoelectronic trend deviates from those predicted by others and shows inexplicable leaps.

\begin{figure*}[ht]
\includegraphics[width=0.45\textwidth, angle=-90]{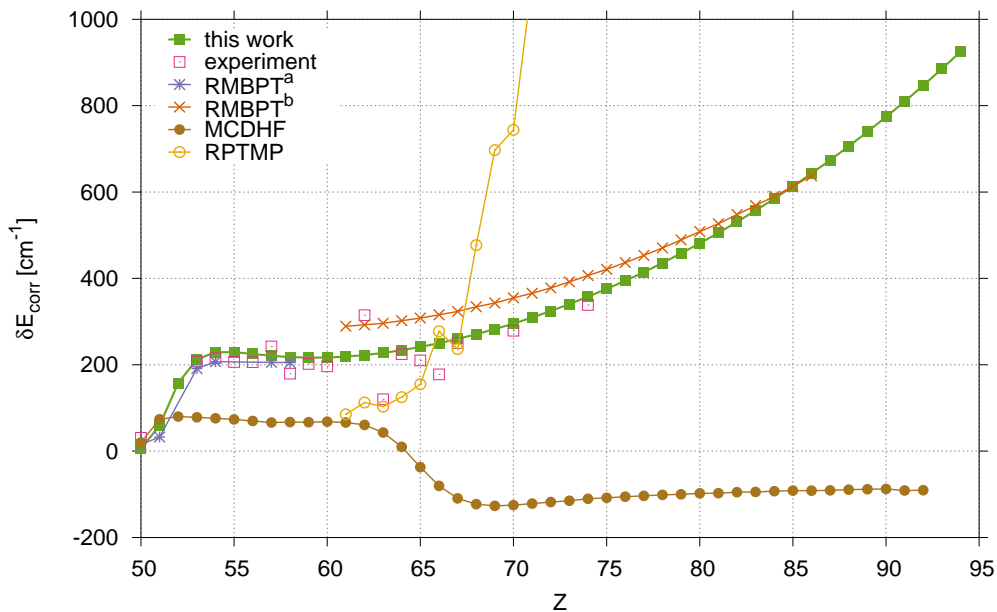}
\caption{\label{fig:corr} Estimated absolute contributions from core-valence correlation to the $4f~^2\mathrm{F}^o$ fine structure separation ($\delta E_\mathrm{corr} = E^\mathrm{method}-E^\mathrm{FCV}_\mathrm{DF}$ where the "method" superscript should be replaced by the corresponding label in the legend) for experiment and other available theory (RMBPT$^a$ \cite{safronova2003}, RMBPT$^b$ \cite{safronova2003} (from \cite{ivanova2011}), MCDHF \cite{ding2012} and RPTMP \cite{ivanova2009,ivanova2011}).}
\end{figure*}

To analyze the different theoretical methods and experiments further, we plot the contribution to the fine structure separation due to electron correlation in Fig. \ref{fig:corr}. This is defined as the best available value (theoretical or experimental) from which the Dirac-Fock value is been subtracted. This again reveals a good agreement between our results and experiment, in terms of the individual data points and in the isoelectronic trend. The RMBPT results also agree well with both our results and the experimental values. Comparing this plot with Fig. \ref{fig:obscalc} one can conclude that the earlier MCDHF \cite{ding2012} calculation lacks a major bulk of electron correlation necessary to reach a fine structure separation close to experimental results. The irregular trend along the sequence mentioned above, especially the dip in energy around $Z=65$, remains unexplained.

\subsection{Magnetic-dipole transition probabilities}
\begin{table*}
\caption{\label{tab:trans} Wavelengths in vacuum ($\lambda_{vac}$), transition rates ($A$), weighted oscillator strengths ($gf$) and line strengths ({$ S$}) of the $4f~^2\mathrm{F}_{5/2}^o~-~^2\mathrm{F}_{7/2}^o$ magnetic-dipole (M1) transition of Ag-like ions between $Z=50$ and $94$ from the large-scale FCV calculation. Note that the $J=7/2$ is the lowest level up to and including $Z=52$, then the two levels cross and and $J=5/2$ becomes the lower of the two from $Z\ge 53$. Numbers in square brackets denotes powers of ten.}
%\footnotesize
\begin{tabular*}{\textwidth}{@{\extracolsep{\fill}}ccccccccccc}
\hline \hline
$Z$ &\multicolumn{1}{c}{$\lambda_{vac}$ (\AA)} &\multicolumn{1}{c}{$A$ (s$^{-1}$)} &\multicolumn{1}{c}{$gf$} &\multicolumn{1}{c}{$S$}  & & 
$Z$ &\multicolumn{1}{c}{$\lambda_{vac}$ (\AA)} &\multicolumn{1}{c}{$A$ (s$^{-1}$)} &\multicolumn{1}{c}{$gf$} &\multicolumn{1}{c}{$S$} \\ \cline{1-5}\cline{7-11}
50 &1.176[$+6$]   &9.463[$-6$]   &1.178[$-8$]   &3.428 & & 73   &3.733[$+3$] &2.216[$+2$] &3.703[$-6$]   &3.419\\
51 &7.378[$+5$]   &3.837[$-5$]   &1.879[$-8$]   &3.428 & & 74   &3.376[$+3$] &2.995[$+2$] &4.094[$-6$]   &3.418\\
52 &2.459[$+6$]   &1.036[$-6$]   &5.638[$-9$]   &3.428 & & 75   &3.063[$+3$] &4.011[$+2$] &4.513[$-6$]   &3.418\\
53 &4.870[$+5$]   &1.000[$-4$]   &2.846[$-8$]   &3.428 & & 76   &2.786[$+3$] &5.326[$+2$] &4.960[$-6$]   &3.417\\
54 &1.774[$+5$]   &2.070[$-3$]   &7.814[$-8$]   &3.428 & & 77   &2.542[$+3$] &7.017[$+2$] &5.436[$-6$]   &3.417\\
55 &9.902[$+4$]   &1.190[$-2$]   &1.400[$-7$]   &3.427 & & 78   &2.324[$+3$] &9.175[$+2$] &5.944[$-6$]   &3.416\\
56 &6.514[$+4$]   &4.180[$-2$]   &2.127[$-7$]   &3.427 & & 79   &2.130[$+3$] &1.191[$+3$] &6.484[$-6$]   &3.415\\
57 &4.676[$+4$]   &1.130[$-1$]   &2.963[$-7$]   &3.426 & & 80   &1.957[$+3$] &1.537[$+3$] &7.057[$-6$]   &3.415\\
58 &3.543[$+4$]   &2.598[$-1$]   &3.910[$-7$]   &3.426 & & 81   &1.801[$+3$] &1.970[$+3$] &7.666[$-6$]   &3.414\\
59 &2.784[$+4$]   &5.350[$-1$]   &4.975[$-7$]   &3.425 & & 82   &1.661[$+3$] &2.511[$+3$] &8.310[$-6$]   &3.414\\
60 &2.247[$+4$]   &1.018[$+0$]   &6.164[$-7$]   &3.425 & & 83   &1.535[$+3$] &3.182[$+3$] &8.992[$-6$]   &3.413\\
61 &1.850[$+4$]   &1.824[$+0$]   &7.486[$-7$]   &3.425 & & 84   &1.421[$+3$] &4.012[$+3$] &9.712[$-6$]   &3.412\\
62 &1.547[$+4$]   &3.117[$+0$]   &8.949[$-7$]   &3.424 & & 85   &1.317[$+3$] &5.032[$+3$] &1.047[$-5$]   &3.412\\
63 &1.311[$+4$]   &5.124[$+0$]   &1.056[$-6$]   &3.424 & & 86   &1.223[$+3$] &6.283[$+3$] &1.128[$-5$]   &3.411\\
64 &1.123[$+4$]   &8.161[$+0$]   &1.233[$-6$]   &3.423 & & 87   &1.138[$+3$] &7.807[$+3$] &1.212[$-5$]   &3.410\\
65 &9.698[$+3$]   &1.265[$+1$]   &1.427[$-6$]   &3.423 & & 88   &1.060[$+3$] &9.660[$+3$] &1.301[$-5$]   &3.410\\
66 &8.444[$+3$]   &1.916[$+1$]   &1.639[$-6$]   &3.422 & & 89   &9.884[$+2$]&1.190[$+4$] &1.395[$-5$]   &3.409\\
67 &7.402[$+3$]   &2.845[$+1$]   &1.869[$-6$]   &3.422 & & 90   &9.231[$+2$]&1.461[$+4$] &1.493[$-5$]   &3.408\\
68 &6.528[$+3$]   &4.148[$+1$]   &2.120[$-6$]   &3.421 & & 91   &8.632[$+2$]&1.786[$+4$] &1.596[$-5$]   &3.407\\
69 &5.787[$+3$]   &5.952[$+1$]   &2.391[$-6$]   &3.421 & & 92   &8.082[$+2$]&2.176[$+4$] &1.705[$-5$]   &3.407\\
70 &5.155[$+3$]   &8.420[$+1$]   &2.683[$-6$]   &3.420 & & 93   &7.576[$+2$]&2.641[$+4$] &1.818[$-5$]   &3.406\\
71 &4.612[$+3$]   &1.176[$+2$]   &2.999[$-6$]   &3.420 & & 94   &7.110[$+2$]&3.195[$+4$] &1.937[$-5$]   &3.405\\
72 &4.142[$+3$]   &1.623[$+2$]   &3.339[$-6$]   &3.419 & &      &     &               &                &              \\
\hline \hline
\end{tabular*}
\end{table*}
The calculation of the magnetic-dipole (M1) transition rate is almost trivial once the correct transition energy has been found, since the M1 operator is independent of the radial part of the wavefunction. In Tab. \ref{tab:trans} the (vacuum) wavelength of the transition is presented along the isoelectronic sequence, together with the corresponding rates, weighted oscillator strengths and line strengths. The simplicity of the M1 transition rate is reflected in the almost constant behavior of the line strength (which is independent of the transition energy). The small decrease seen with increasing $Z$ is due to the CSF composition of the wavefunctions of the involved states, via interaction with other $LS$-terms than $^2\mathrm{F}^o$. This effect is however small since the ground term $^2\mathrm{F}^o$ is well-isolated in energy for most of the ions in the sequence.

\section{Conclusions}
In this work we have presented a systematic MCDHF study of the $4f~^2\mathrm{F}^o$ fine structure separation and the involved magnetic-dipole transition for Ag-like ions with nuclear charges $Z=50-94$. Special attention has been payed to core-valence effects with deep core subshells and it was shown that core-valence correlation with $3d$, rather than $4d$, is the dominant contributor for intermediate and highly charged ions. The underlying reason for this could be an interesting case for further studies. Our large-scale MCDHF calculations include correlation effects from Coulomb- and (frequency independent) Breit interaction, as well as corrections due to dominant QED effects. The accuracy of the $^2\mathrm{F}^o$ fine structure separation is carefully analyzed through systematic studies of convergence trends as the active set of virtual Dirac-orbitals, used to construct the many-body basis, is increased. This is augmented by studies of the smoothness of different properties along the isoelectronic sequence. Furthermore, a good agreement with experiments, of which some are very recent EBIT measurements \cite{zhao2014, fei2012}, and other reliable theoretical results, finally leads us to conclude that our method provides accurate data for the $^2\mathrm{F}^o$ levels of Ag-like ions. Transition rates, weighted oscillator strengths and line strengths of the magnetic-dipole transition between these two fine structure levels have been calculated and tabulated. These data should be accurate since the M1 operator is not dependent on the radial part of the wavefunctions, and the transitions energies are accurately predicted. 

\section{Acknowledgments}
This work was supported by the National Natural Science Foundation of China under project no.~11074049, and by the Shanghai Leading Academic Discipline Project B107. We also gratefully acknowledge support from the 
Swedish Research Council (Vetenskapsr\aa det) and the Swedish Institute under the Visby-programme. JG and WL would like to especially thank the Nordic Centre at Fudan University for supporting an exchange between Lund and Fudan. Finally the authors would like to thank Gordon Berry, J\"orgen Ekman and Per J\"onsson for valuable discussions.

\bibliography{refs}

\begin{thebibliography}{29}
\expandafter\ifx\csname natexlab\endcsname\relax\def\natexlab#1{#1}\fi
\expandafter\ifx\csname bibnamefont\endcsname\relax
  \def\bibnamefont#1{#1}\fi
\expandafter\ifx\csname bibfnamefont\endcsname\relax
  \def\bibfnamefont#1{#1}\fi
\expandafter\ifx\csname citenamefont\endcsname\relax
  \def\citenamefont#1{#1}\fi
\expandafter\ifx\csname url\endcsname\relax
  \def\url#1{\texttt{#1}}\fi
\expandafter\ifx\csname urlprefix\endcsname\relax\def\urlprefix{URL }\fi
\providecommand{\bibinfo}[2]{#2}
\providecommand{\eprint}[2][]{\url{#2}}

\bibitem[{\citenamefont{Edl{\'e}n}(1943)}]{edlen1943}
\bibinfo{author}{\bibfnamefont{B.}~\bibnamefont{Edl{\'e}n}},
  \bibinfo{journal}{Zeitschrift f\"ur Astrophysik}
  \textbf{\bibinfo{volume}{22}}, \bibinfo{pages}{30} (\bibinfo{year}{1943}).

\bibitem[{\citenamefont{Edl{\'e}n}(1945)}]{edlen1945}
\bibinfo{author}{\bibfnamefont{B.}~\bibnamefont{Edl{\'e}n}},
  \bibinfo{journal}{Monthly Notices of the Royal Astronomical Society}
  \textbf{\bibinfo{volume}{105}}, \bibinfo{pages}{323} (\bibinfo{year}{1945}).

\bibitem[{\citenamefont{Suckewer and Hinnov}(1978)}]{suckewer1978}
\bibinfo{author}{\bibfnamefont{S.}~\bibnamefont{Suckewer}} \bibnamefont{and}
  \bibinfo{author}{\bibfnamefont{E.}~\bibnamefont{Hinnov}},
  \bibinfo{journal}{Phys. Rev. Lett.} \textbf{\bibinfo{volume}{41}},
  \bibinfo{pages}{756} (\bibinfo{year}{1978}),
  \urlprefix\url{http://link.aps.org/doi/10.1103/PhysRevLett.41.756}.

\bibitem[{\citenamefont{Sugar and Kaufman}(1980)}]{sugar1980a}
\bibinfo{author}{\bibfnamefont{J.}~\bibnamefont{Sugar}} \bibnamefont{and}
  \bibinfo{author}{\bibfnamefont{V.}~\bibnamefont{Kaufman}},
  \bibinfo{journal}{Phys. Rev. A} \textbf{\bibinfo{volume}{21}},
  \bibinfo{pages}{2096} (\bibinfo{year}{1980}),
  \urlprefix\url{http://link.aps.org/doi/10.1103/PhysRevA.21.2096}.

\bibitem[{\citenamefont{Fei et~al.}(2012)\citenamefont{Fei, Zhao, Shi, Xiao,
  Qiu, Grumer, Andersson, Brage, Hutton, and Zou}}]{fei2012}
\bibinfo{author}{\bibfnamefont{Z.}~\bibnamefont{Fei}},
  \bibinfo{author}{\bibfnamefont{R.}~\bibnamefont{Zhao}},
  \bibinfo{author}{\bibfnamefont{Z.}~\bibnamefont{Shi}},
  \bibinfo{author}{\bibfnamefont{J.}~\bibnamefont{Xiao}},
  \bibinfo{author}{\bibfnamefont{M.}~\bibnamefont{Qiu}},
  \bibinfo{author}{\bibfnamefont{J.}~\bibnamefont{Grumer}},
  \bibinfo{author}{\bibfnamefont{M.}~\bibnamefont{Andersson}},
  \bibinfo{author}{\bibfnamefont{T.}~\bibnamefont{Brage}},
  \bibinfo{author}{\bibfnamefont{R.}~\bibnamefont{Hutton}}, \bibnamefont{and}
  \bibinfo{author}{\bibfnamefont{Y.}~\bibnamefont{Zou}},
  \bibinfo{journal}{Physical Review A} \textbf{\bibinfo{volume}{86}},
  \bibinfo{pages}{062501} (\bibinfo{year}{2012}).

\bibitem[{\citenamefont{Zhao}(2014)}]{zhao2014}
\bibinfo{author}{\bibfnamefont{R.}~\bibnamefont{Zhao}},
  \bibinfo{journal}{private communication}  (\bibinfo{year}{2014}).

\bibitem[{\citenamefont{Morita et~al.}(2013)\citenamefont{Morita, Dong, Goto,
  Kato, Murakami, Sakaue, Hasuo, Koike, Nakamura, Oishi et~al.}}]{morita2013}
\bibinfo{author}{\bibfnamefont{S.}~\bibnamefont{Morita}},
  \bibinfo{author}{\bibfnamefont{C.~F.} \bibnamefont{Dong}},
  \bibinfo{author}{\bibfnamefont{M.}~\bibnamefont{Goto}},
  \bibinfo{author}{\bibfnamefont{D.}~\bibnamefont{Kato}},
  \bibinfo{author}{\bibfnamefont{I.}~\bibnamefont{Murakami}},
  \bibinfo{author}{\bibfnamefont{H.~A.} \bibnamefont{Sakaue}},
  \bibinfo{author}{\bibfnamefont{M.}~\bibnamefont{Hasuo}},
  \bibinfo{author}{\bibfnamefont{F.}~\bibnamefont{Koike}},
  \bibinfo{author}{\bibfnamefont{N.}~\bibnamefont{Nakamura}},
  \bibinfo{author}{\bibfnamefont{T.}~\bibnamefont{Oishi}},
  \bibnamefont{et~al.}, \bibinfo{journal}{AIP Conference Proceedings}
  \textbf{\bibinfo{volume}{1545}}, \bibinfo{pages}{143} (\bibinfo{year}{2013}),
  \urlprefix\url{http://scitation.aip.org/content/aip/proceeding/aipcp/10.1063/1.4815848}.

\bibitem[{\citenamefont{Martinson and Jup{\'e}n}(2001)}]{martinson2001}
\bibinfo{author}{\bibfnamefont{I.}~\bibnamefont{Martinson}} \bibnamefont{and}
  \bibinfo{author}{\bibfnamefont{C.}~\bibnamefont{Jup{\'e}n}},
  \bibinfo{journal}{Journal of the Chinese Chemical Society}
  \textbf{\bibinfo{volume}{48}}, \bibinfo{pages}{469} (\bibinfo{year}{2001}),
  ISSN \bibinfo{issn}{2192-6549},
  \urlprefix\url{http://dx.doi.org/10.1002/jccs.200100070}.

\bibitem[{\citenamefont{Lapierre et~al.}(2005)\citenamefont{Lapierre,
  Jentschura, L{\'o}pez-Urrutia, Braun, Brenner, Bruhns, Fischer,
  Mart{\'\i}nez, Harman, Johnson et~al.}}]{lapierre2005}
\bibinfo{author}{\bibfnamefont{A.}~\bibnamefont{Lapierre}},
  \bibinfo{author}{\bibfnamefont{U.}~\bibnamefont{Jentschura}},
  \bibinfo{author}{\bibfnamefont{J.~C.} \bibnamefont{L{\'o}pez-Urrutia}},
  \bibinfo{author}{\bibfnamefont{J.}~\bibnamefont{Braun}},
  \bibinfo{author}{\bibfnamefont{G.}~\bibnamefont{Brenner}},
  \bibinfo{author}{\bibfnamefont{H.}~\bibnamefont{Bruhns}},
  \bibinfo{author}{\bibfnamefont{D.}~\bibnamefont{Fischer}},
  \bibinfo{author}{\bibfnamefont{A.~G.} \bibnamefont{Mart{\'\i}nez}},
  \bibinfo{author}{\bibfnamefont{Z.}~\bibnamefont{Harman}},
  \bibinfo{author}{\bibfnamefont{W.}~\bibnamefont{Johnson}},
  \bibnamefont{et~al.}, \bibinfo{journal}{Phys. Rev. Lett.}
  \textbf{\bibinfo{volume}{95}}, \bibinfo{pages}{183001}
  (\bibinfo{year}{2005}),
  \urlprefix\url{http://link.aps.org/doi/10.1103/PhysRevLett.95.183001}.

\bibitem[{\citenamefont{Tr\"abert et~al.}(2007)\citenamefont{Tr\"abert,
  Beiersdorfer, and Brown}}]{trabert2007}
\bibinfo{author}{\bibfnamefont{E.}~\bibnamefont{Tr\"abert}},
  \bibinfo{author}{\bibfnamefont{P.}~\bibnamefont{Beiersdorfer}},
  \bibnamefont{and} \bibinfo{author}{\bibfnamefont{G.~V.} \bibnamefont{Brown}},
  \bibinfo{journal}{Phys. Rev. Lett.} \textbf{\bibinfo{volume}{98}},
  \bibinfo{pages}{263001} (\bibinfo{year}{2007}),
  \urlprefix\url{http://link.aps.org/doi/10.1103/PhysRevLett.98.263001}.

\bibitem[{\citenamefont{Ralchenko}(2007)}]{ralchenko2007}
\bibinfo{author}{\bibfnamefont{Y.}~\bibnamefont{Ralchenko}},
  \bibinfo{journal}{Journal of Physics B: Atomic, Molecular and Optical
  Physics} \textbf{\bibinfo{volume}{40}}, \bibinfo{pages}{F175}
  (\bibinfo{year}{2007}),
  \urlprefix\url{http://stacks.iop.org/0953-4075/40/i=11/a=F01}.

\bibitem[{\citenamefont{Safronova et~al.}(2003)\citenamefont{Safronova,
  Savukov, Safronova, and Johnson}}]{safronova2003}
\bibinfo{author}{\bibfnamefont{U.~I.} \bibnamefont{Safronova}},
  \bibinfo{author}{\bibfnamefont{I.~M.} \bibnamefont{Savukov}},
  \bibinfo{author}{\bibfnamefont{M.~S.} \bibnamefont{Safronova}},
  \bibnamefont{and} \bibinfo{author}{\bibfnamefont{W.~R.}
  \bibnamefont{Johnson}}, \bibinfo{journal}{Phys. Rev. A}
  \textbf{\bibinfo{volume}{68}}, \bibinfo{pages}{062505}
  (\bibinfo{year}{2003}),
  \urlprefix\url{http://link.aps.org/doi/10.1103/PhysRevA.68.062505}.

\bibitem[{\citenamefont{Ivanova}(2011)}]{ivanova2011}
\bibinfo{author}{\bibfnamefont{E.}~\bibnamefont{Ivanova}},
  \bibinfo{journal}{Atomic Data and Nuclear Data Tables}
  \textbf{\bibinfo{volume}{97}}, \bibinfo{pages}{1} (\bibinfo{year}{2011}).

\bibitem[{\citenamefont{Ivanova}(2009)}]{ivanova2009}
\bibinfo{author}{\bibfnamefont{E.}~\bibnamefont{Ivanova}},
  \bibinfo{journal}{Atomic Data and Nuclear Data Tables}
  \textbf{\bibinfo{volume}{95}} (\bibinfo{year}{2009}).

\bibitem[{\citenamefont{Ding et~al.}(2012)\citenamefont{Ding, Koike, Murakami,
  Kato, Sakaue, Dong, and Nakamura}}]{ding2012}
\bibinfo{author}{\bibfnamefont{X.-B.} \bibnamefont{Ding}},
  \bibinfo{author}{\bibfnamefont{F.}~\bibnamefont{Koike}},
  \bibinfo{author}{\bibfnamefont{I.}~\bibnamefont{Murakami}},
  \bibinfo{author}{\bibfnamefont{D.}~\bibnamefont{Kato}},
  \bibinfo{author}{\bibfnamefont{H.~A.} \bibnamefont{Sakaue}},
  \bibinfo{author}{\bibfnamefont{C.-Z.} \bibnamefont{Dong}}, \bibnamefont{and}
  \bibinfo{author}{\bibfnamefont{N.}~\bibnamefont{Nakamura}},
  \bibinfo{journal}{Journal of Physics B: Atomic, Molecular and Optical
  Physics} \textbf{\bibinfo{volume}{45}}, \bibinfo{pages}{035003}
  (\bibinfo{year}{2012}).

\bibitem[{\citenamefont{{J{\"o}nsson} et~al.}(2013)\citenamefont{{J{\"o}nsson},
  {Gaigalas}, {Biero{\'n}}, {Fischer}, and {Grant}}}]{jonsson2013}
\bibinfo{author}{\bibfnamefont{P.}~\bibnamefont{{J{\"o}nsson}}},
  \bibinfo{author}{\bibfnamefont{G.}~\bibnamefont{{Gaigalas}}},
  \bibinfo{author}{\bibfnamefont{J.}~\bibnamefont{{Biero{\'n}}}},
  \bibinfo{author}{\bibfnamefont{C.~F.} \bibnamefont{{Fischer}}},
  \bibnamefont{and} \bibinfo{author}{\bibfnamefont{I.~P.}
  \bibnamefont{{Grant}}}, \bibinfo{journal}{Computer Physics Communications}
  \textbf{\bibinfo{volume}{184}}, \bibinfo{pages}{2197 }
  (\bibinfo{year}{2013}), ISSN \bibinfo{issn}{0010-4655},
  \urlprefix\url{http://www.sciencedirect.com/science/article/pii/S0010465513000738}.

\bibitem[{\citenamefont{Dyall et~al.}(1989)\citenamefont{Dyall, Grant, Johnson,
  Parpia, and Plummer}}]{dyall1989}
\bibinfo{author}{\bibfnamefont{K.}~\bibnamefont{Dyall}},
  \bibinfo{author}{\bibfnamefont{I.}~\bibnamefont{Grant}},
  \bibinfo{author}{\bibfnamefont{C.}~\bibnamefont{Johnson}},
  \bibinfo{author}{\bibfnamefont{F.}~\bibnamefont{Parpia}}, \bibnamefont{and}
  \bibinfo{author}{\bibfnamefont{E.}~\bibnamefont{Plummer}},
  \bibinfo{journal}{Computer Physics Communications}
  \textbf{\bibinfo{volume}{55}}, \bibinfo{pages}{425} (\bibinfo{year}{1989}).

\bibitem[{\citenamefont{Parpia et~al.}(1996)\citenamefont{Parpia, Fischer, and
  Grant}}]{parpia1996}
\bibinfo{author}{\bibfnamefont{F.~A.} \bibnamefont{Parpia}},
  \bibinfo{author}{\bibfnamefont{C.~F.} \bibnamefont{Fischer}},
  \bibnamefont{and} \bibinfo{author}{\bibfnamefont{I.~P.} \bibnamefont{Grant}},
  \bibinfo{journal}{Computer physics communications}
  \textbf{\bibinfo{volume}{94}}, \bibinfo{pages}{249} (\bibinfo{year}{1996}).

\bibitem[{\citenamefont{Grant}(2006)}]{grant2006}
\bibinfo{author}{\bibfnamefont{I.~P.} \bibnamefont{Grant}},
  \emph{\bibinfo{title}{Relativistic Quantum Theory of Atoms and Molecules:
  Theory and Computation (Springer Series on Atomic, Optical, and Plasma
  Physics)}} (\bibinfo{publisher}{Springer-Verlag New York, Inc.},
  \bibinfo{address}{Secaucus, NJ, USA}, \bibinfo{year}{2006}), ISBN
  \bibinfo{isbn}{0387346716}.

\bibitem[{\citenamefont{Froese~Fischer
  et~al.}(1997)\citenamefont{Froese~Fischer, Brage, and
  J{\"o}nsson}}]{froese1997book}
\bibinfo{author}{\bibfnamefont{C.}~\bibnamefont{Froese~Fischer}},
  \bibinfo{author}{\bibfnamefont{T.}~\bibnamefont{Brage}}, \bibnamefont{and}
  \bibinfo{author}{\bibfnamefont{P.}~\bibnamefont{J{\"o}nsson}},
  \emph{\bibinfo{title}{Computational atomic structure: an MCHF approach}}
  (\bibinfo{publisher}{Inst. of Physics Publishing, Bristol},
  \bibinfo{year}{1997}).

\bibitem[{\citenamefont{Fullerton and Rinker}(1976)}]{fullerton1976}
\bibinfo{author}{\bibfnamefont{L.~W.} \bibnamefont{Fullerton}}
  \bibnamefont{and} \bibinfo{author}{\bibfnamefont{G.~A.}
  \bibnamefont{Rinker}}, \bibinfo{journal}{Phys. Rev. A}
  \textbf{\bibinfo{volume}{13}}, \bibinfo{pages}{1283} (\bibinfo{year}{1976}),
  \urlprefix\url{http://link.aps.org/doi/10.1103/PhysRevA.13.1283}.

\bibitem[{\citenamefont{Mohr}(1983)}]{mohr1983}
\bibinfo{author}{\bibfnamefont{P.~J.} \bibnamefont{Mohr}},
  \bibinfo{journal}{Atomic Data and Nuclear Data Tables}
  \textbf{\bibinfo{volume}{29}}, \bibinfo{pages}{453 } (\bibinfo{year}{1983}),
  ISSN \bibinfo{issn}{0092-640X},
  \urlprefix\url{http://www.sciencedirect.com/science/article/pii/S0092640X83800023}.

\bibitem[{\citenamefont{Moore}(1971)}]{moore1958}
\bibinfo{author}{\bibfnamefont{C.}~\bibnamefont{Moore}},
  \bibinfo{journal}{Stand. Ref. Data Ser., Nat. Bur. Stand.(US)}
  \textbf{\bibinfo{volume}{35}} (\bibinfo{year}{1971}).

\bibitem[{\citenamefont{Kramida et~al.}(2012)\citenamefont{Kramida, Ralchenko,
  Reader, and {NIST ASD Team}}}]{NIST}
\bibinfo{author}{\bibfnamefont{A.}~\bibnamefont{Kramida}},
  \bibinfo{author}{\bibfnamefont{Y.}~\bibnamefont{Ralchenko}},
  \bibinfo{author}{\bibfnamefont{J.}~\bibnamefont{Reader}}, \bibnamefont{and}
  \bibinfo{author}{\bibnamefont{{NIST ASD Team}}}, \bibinfo{journal}{NIST
  Atomic Spectra Database (ver. 5.0) [Online]} p.
  \bibinfo{pages}{http://www.nist.gov/pml/data/asd.cfm} (\bibinfo{year}{2012}),
  \urlprefix\url{[Online]http://physics.nist.gov/asd}.

\bibitem[{\citenamefont{Kaufman and Sugar}(1981)}]{kaufman1981a}
\bibinfo{author}{\bibfnamefont{V.}~\bibnamefont{Kaufman}} \bibnamefont{and}
  \bibinfo{author}{\bibfnamefont{J.}~\bibnamefont{Sugar}},
  \bibinfo{journal}{Physica Scripta} \textbf{\bibinfo{volume}{24}},
  \bibinfo{pages}{738} (\bibinfo{year}{1981}).

\bibitem[{\citenamefont{Larsson et~al.}(1995)\citenamefont{Larsson, Gonzalez,
  Hallin, Heijkensk\"old, Hutton, Langereis, Nystr\"om, O'Sullivan, and
  W\"annstr\"om}}]{larsson1995}
\bibinfo{author}{\bibfnamefont{M.~O.} \bibnamefont{Larsson}},
  \bibinfo{author}{\bibfnamefont{A.~M.} \bibnamefont{Gonzalez}},
  \bibinfo{author}{\bibfnamefont{R.}~\bibnamefont{Hallin}},
  \bibinfo{author}{\bibfnamefont{F.}~\bibnamefont{Heijkensk\"old}},
  \bibinfo{author}{\bibfnamefont{R.}~\bibnamefont{Hutton}},
  \bibinfo{author}{\bibfnamefont{A.}~\bibnamefont{Langereis}},
  \bibinfo{author}{\bibfnamefont{B.}~\bibnamefont{Nystr\"om}},
  \bibinfo{author}{\bibfnamefont{G.}~\bibnamefont{O'Sullivan}},
  \bibnamefont{and}
  \bibinfo{author}{\bibfnamefont{A.}~\bibnamefont{W\"annstr\"om}},
  \bibinfo{journal}{Physica Scripta} \textbf{\bibinfo{volume}{51}},
  \bibinfo{pages}{69} (\bibinfo{year}{1995}),
  \urlprefix\url{http://stacks.iop.org/1402-4896/51/i=1/a=011}.

\bibitem[{\citenamefont{Tauheed and Joshi}(2005)}]{tauheed2005}
\bibinfo{author}{\bibfnamefont{A.}~\bibnamefont{Tauheed}} \bibnamefont{and}
  \bibinfo{author}{\bibfnamefont{Y.}~\bibnamefont{Joshi}},
  \bibinfo{journal}{Physica Scripta} \textbf{\bibinfo{volume}{72}},
  \bibinfo{pages}{385} (\bibinfo{year}{2005}).

\bibitem[{\citenamefont{Churilov and Joshi}(2000)}]{churilov2000}
\bibinfo{author}{\bibfnamefont{S.}~\bibnamefont{Churilov}} \bibnamefont{and}
  \bibinfo{author}{\bibfnamefont{Y.}~\bibnamefont{Joshi}},
  \bibinfo{journal}{Physica Scripta} \textbf{\bibinfo{volume}{62}},
  \bibinfo{pages}{282} (\bibinfo{year}{2000}).

\bibitem[{\citenamefont{Sugar and Kaufman}(1981)}]{sugar1981b}
\bibinfo{author}{\bibfnamefont{J.}~\bibnamefont{Sugar}} \bibnamefont{and}
  \bibinfo{author}{\bibfnamefont{V.}~\bibnamefont{Kaufman}},
  \bibinfo{journal}{Physica Scripta} \textbf{\bibinfo{volume}{24}},
  \bibinfo{pages}{742} (\bibinfo{year}{1981}).

\end{thebibliography}

\end{document}